\pdfoutput=1
\documentclass[12pt,a4paper,titlepage]{article}
\usepackage[utf8]{inputenc}
\usepackage[top=50pt,bottom=50pt,left=68pt,right=66pt]{geometry}
\usepackage{amsmath}
\usepackage{booktabs}
\usepackage{amssymb}
\usepackage[title]{appendix}
\usepackage{mathrsfs}
\usepackage{float}
\usepackage{multirow}
\usepackage{graphicx,caption,subcaption}
\usepackage[space]{grffile}

\begin{document}
\renewcommand{\arraystretch}{1.3}

\makeatletter
\def\@hangfrom#1{\setbox\@tempboxa\hbox{{#1}}%
      \hangindent 0pt
      \noindent\box\@tempboxa}
\makeatother


\def\un#1{\relax\ifmmode\@@underline#1\else
        $\@@underline{\hbox{#1}}$\relax\fi}


\let\under=\unt                 
\let\ced=\ce                    
\let\du=\du                     
\let\um=\Hu                     
\let\sll=\lp                    
\let\Sll=\Lp                    
\let\slo=\os                    
\let\Slo=\Os                    
\let\tie=\ta                    
\let\br=\ub                     


\def\a{\alpha}
\def\b{\beta}
\def\c{\chi}
\def\d{\delta}
\def\e{\epsilon}
\def\f{\phi}
\def\g{\gamma}
\def\h{\eta}
\def\i{\iota}
\def\j{\psi}
\def\k{\kappa}
\def\l{\lambda}
\def\m{\mu}
\def\n{\nu}
\def\o{\omega}
\def\p{\pi}
\def\q{\theta}
\def\r{\rho}
\def\s{\sigma}
\def\t{\tau}
\def\u{\upsilon}
\def\x{\xi}
\def\z{\zeta}
\def\D{\Delta}
\def\F{\Phi}
\def\G{\Gamma}
\def\J{\Psi}
\def\L{\Lambda}
\def\O{\Omega}
\def\P{\Pi}
\def\Q{\Theta}
\def\S{\Sigma}
\def\U{\Upsilon}
\def\X{\Xi}


\def\ve{\varepsilon}
\def\vf{\varphi}
\def\vr{\varrho}
\def\vs{\varsigma}
\def\vq{\vartheta}


\def\ca{{\cal A}}
\def\cb{{\cal B}}
\def\cc{{\cal C}}
\def\cd{{\cal D}}
\def\ce{{\cal E}}
\def\cf{{\cal F}}
\def\cg{{\cal G}}
\def\ch{{\cal H}}
\def\ci{{\cal I}}
\def\cj{{\cal J}}
\def\ck{{\cal K}}
\def\cl{{\cal L}}
\def\cm{{\cal M}}
\def\cn{{\cal N}}
\def\co{{\cal O}}
\def\cp{{\cal P}}
\def\cq{{\cal Q}}
\def\car{{\cal R}}
\def\cs{{\cal S}}
\def\ct{{\cal T}}
\def\cu{{\cal U}}
\def\cv{{\cal V}}
\def\cw{{\cal W}}
\def\cx{{\cal X}}
\def\cy{{\cal Y}}
\def\cz{{\cal Z}}


\def\Sc#1{{\hbox{\sc #1}}}      
\def\Sf#1{{\hbox{\sf #1}}}      



\def\slpa{\slash{\pa}}                            
\def\slin{\SLLash{\in}}                                   
\def\bo{{\raise-.3ex\hbox{\large$\Box$}}}               
\def\cbo{\Sc [}                                         
\def\pa{\partial}                                       
\def\de{\nabla}                                         
\def\dell{\bigtriangledown}                             
\def\su{\sum}                                           
\def\pr{\prod}                                          
\def\iff{\leftrightarrow}                               
\def\conj{{\hbox{\large *}}}                            
\def\ltap{\raisebox{-.4ex}{\rlap{$\sim$}} \raisebox{.4ex}{$<$}}   
\def\gtap{\raisebox{-.4ex}{\rlap{$\sim$}} \raisebox{.4ex}{$>$}}   
\def\TH{{\raise.2ex\hbox{$\displaystyle \bigodot$}\mskip-4.7mu \llap H \;}}
\def\face{{\raise.2ex\hbox{$\displaystyle \bigodot$}\mskip-2.2mu \llap {$\ddot
        \smile$}}}                                      
\def\dg{\sp\dagger}                                     
\def\ddg{\sp\ddagger}                                   

\font\tenex=cmex10 scaled 1200


\def\sp#1{{}^{#1}}                              
\def\sb#1{{}_{#1}}                              
\def\oldsl#1{\rlap/#1}                          
\def\slash#1{\rlap{\hbox{$\mskip 1 mu /$}}#1}      
\def\Slash#1{\rlap{\hbox{$\mskip 3 mu /$}}#1}      
\def\SLash#1{\rlap{\hbox{$\mskip 4.5 mu /$}}#1}    
\def\SLLash#1{\rlap{\hbox{$\mskip 6 mu /$}}#1}      
\def\PMMM#1{\rlap{\hbox{$\mskip 2 mu | $}}#1}   %
\def\PMM#1{\rlap{\hbox{$\mskip 4 mu ~ \mid $}}#1}       %
\def\Tilde#1{\widetilde{#1}}                    
\def\Hat#1{\widehat{#1}}                        
\def\Bar#1{\overline{#1}}                       
\def\sbar#1{\stackrel{*}{\Bar{#1}}}             
\def\bra#1{\left\langle #1\right|}              
\def\ket#1{\left| #1\right\rangle}              
\def\VEV#1{\left\langle #1\right\rangle}        
\def\abs#1{\left| #1\right|}                    
\def\leftrightarrowfill{$\mathsurround=0pt \mathord\leftarrow \mkern-6mu
        \cleaders\hbox{$\mkern-2mu \mathord- \mkern-2mu$}\hfill
        \mkern-6mu \mathord\rightarrow$}
\def\dvec#1{\vbox{\ialign{##\crcr
        \leftrightarrowfill\crcr\noalign{\kern-1pt\nointerlineskip}
        $\hfil\displaystyle{#1}\hfil$\crcr}}}           
\def\dt#1{{\buildrel {\hbox{\LARGE .}} \over {#1}}}     
\def\dtt#1{{\buildrel \bullet \over {#1}}}              
\def\der#1{{\pa \over \pa {#1}}}                
\def\fder#1{{\d \over \d {#1}}}                 


\def\frac#1#2{{\textstyle{#1\over\vphantom2\smash{\raise.20ex
        \hbox{$\scriptstyle{#2}$}}}}}                   
\def\half{\frac12}                                        
\def\sfrac#1#2{{\vphantom1\smash{\lower.5ex\hbox{\small$#1$}}\over
        \vphantom1\smash{\raise.4ex\hbox{\small$#2$}}}} 
\def\bfrac#1#2{{\vphantom1\smash{\lower.5ex\hbox{$#1$}}\over
        \vphantom1\smash{\raise.3ex\hbox{$#2$}}}}       
\def\afrac#1#2{{\vphantom1\smash{\lower.5ex\hbox{$#1$}}\over#2}}    
\def\partder#1#2{{\partial #1\over\partial #2}}   
\def\parvar#1#2{{\d #1\over \d #2}}               
\def\secder#1#2#3{{\partial^2 #1\over\partial #2 \partial #3}}  
\def\on#1#2{\mathop{\null#2}\limits^{#1}}               
\def\bvec#1{\on\leftarrow{#1}}                  
\def\oover#1{\on\circ{#1}}                              

\def\[{\lfloor{\hskip 0.35pt}\!\!\!\lceil}
\def\]{\rfloor{\hskip 0.35pt}\!\!\!\rceil}
\def\Lag{{\cal L}}
\def\du#1#2{_{#1}{}^{#2}}
\def\ud#1#2{^{#1}{}_{#2}}
\def\dud#1#2#3{_{#1}{}^{#2}{}_{#3}}
\def\udu#1#2#3{^{#1}{}_{#2}{}^{#3}}
\def\calD{{\cal D}}
\def\calM{{\cal M}}

\def\szet{{${\scriptstyle \b}$}}
\def\ulA{{\un A}}
\def\ulM{{\underline M}}
\def\cdm{{\Sc D}_{--}}
\def\cdp{{\Sc D}_{++}}
\def\vTheta{\check\Theta}
\def\fracm#1#2{\hbox{\large{${\frac{{#1}}{{#2}}}$}}}
\def\ha{{\fracmm12}}
\def\tr{{\rm tr}}
\def\Tr{{\rm Tr}}
\def\itrema{$\ddot{\scriptstyle 1}$}
\def\ula{{\underline a}} \def\ulb{{\underline b}} \def\ulc{{\underline c}}
\def\uld{{\underline d}} \def\ule{{\underline e}} \def\ulf{{\underline f}}
\def\ulg{{\underline g}}
\def\items#1{\\ \item{[#1]}}
\def\ul{\underline}
\def\un{\underline}
\def\fracmm#1#2{{{#1}\over{#2}}}
\def\footnotew#1{\footnote{\hsize=6.5in {#1}}}
\def\low#1{{\raise -3pt\hbox{${\hskip 0.75pt}\!_{#1}$}}}

\def\Dot#1{\buildrel{_{_{\hskip 0.01in}\bullet}}\over{#1}}
\def\dt#1{\Dot{#1}}

\def\DDot#1{\buildrel{_{_{\hskip 0.01in}\bullet\bullet}}\over{#1}}
\def\ddt#1{\DDot{#1}}

\def\DDDot#1{\buildrel{_{_{\hskip 0.01in}\bullet\bullet\bullet}}\over{#1}}
\def\dddt#1{\DDDot{#1}}

\def\DDDDot#1{\buildrel{_{_{\hskip 
0.01in}\bullet\bullet\bullet\bullet}}\over{#1}}
\def\ddddt#1{\DDDDot{#1}}

\def\Tilde#1{{\widetilde{#1}}\hskip 0.015in}
\def\Hat#1{\widehat{#1}}


\newskip\humongous \humongous=0pt plus 1000pt minus 1000pt
\def\caja{\mathsurround=0pt}
\def\eqalign#1{\,\vcenter{\openup2\jot \caja
        \ialign{\strut \hfil$\displaystyle{##}$&$
        \displaystyle{{}##}$\hfil\crcr#1\crcr}}\,}
\newif\ifdtup
\def\panorama{\global\dtuptrue \openup2\jot \caja
        \everycr{\noalign{\ifdtup \global\dtupfalse
        \vskip-\lineskiplimit \vskip\normallineskiplimit
        \else \penalty\interdisplaylinepenalty \fi}}}
\def\li#1{\panorama \tabskip=\humongous                         
        \halign to\displaywidth{\hfil$\displaystyle{##}$
        \tabskip=0pt&$\displaystyle{{}##}$\hfil
        \tabskip=\humongous&\llap{$##$}\tabskip=0pt
        \crcr#1\crcr}}
\def\eqalignnotwo#1{\panorama \tabskip=\humongous
        \halign to\displaywidth{\hfil$\displaystyle{##}$
        \tabskip=0pt&$\displaystyle{{}##}$
        \tabskip=0pt&$\displaystyle{{}##}$\hfil
        \tabskip=\humongous&\llap{$##$}\tabskip=0pt
        \crcr#1\crcr}}


\def\eV{\,{\rm eV}}
\def\keV{\,{\rm keV}}
\def\MeV{\,{\rm MeV}}
\def\GeV{\,{\rm GeV}}
\def\TeV{\,{\rm TeV}}
\def\sv{\left<\sigma v\right>}
\def\({\left(}
\def\){\right)}
\def\cm{{\,\rm cm}}
\def\K{{\,\rm K}}
\def\kpc{{\,\rm kpc}}
\def\beq{\begin{equation}}
\def\eeq{\end{equation}}
\def\bea{\begin{eqnarray}}
\def\eea{\end{eqnarray}}


\newcommand{\be}{\begin{equation}}
\newcommand{\ee}{\end{equation}}
\newcommand{\nbe}{\begin{equation*}}
\newcommand{\nee}{\end{equation*}}

\newcommand{\fr}{\frac}
\newcommand{\lb}{\label}

\thispagestyle{empty}

{\hbox to\hsize{
\vbox{\noindent June 2024 \hfill IPMU24-0026} }}

\noindent  

\noindent
\vskip2.0cm
\begin{center}

{\large\bf Starobinsky inflation and Swampland conjectures}
\vglue.2in
{\it In memory of Vladislav Gavrilovich Bagrov}
\vglue.3in
Sergei V. Ketov~${}^{a,b,c,d,\#}$
\vglue.1in

${}^a$~Department of Physics, Tokyo Metropolitan University,\\
1-1 Minami-ohsawa, Hachioji-shi, Tokyo 192-0397, Japan \\
${}^b$~Kavli Institute for the Physics and Mathematics of the Universe,
\\The University of Tokyo Institutes for Advanced Study,\\
  Kashiwa 277-8583, Japan\\
${}^c$~Research School of High-Energy Physics, Tomsk Polytechnic University, \\
Tomsk 634028, Russian Federation\\
${}^d$~Interdisciplinary Research Laboratory, Tomsk State University,\\
Tomsk 634050, Russian Federation\\

\vglue.2in
${}^{\#}$~ketov@tmu.ac.jp
\end{center}

\vglue.4in

\begin{center}
{\Large\bf Abstract}  
\end{center}
\vglue.1in

\noindent This paper is devoted to memory of late Professor V. G. Bagrov, who was my first teacher in theoretical physics. About 45 years
later, theoretical physics has changed and me too. The subject of this paper relates some old ideas in cosmology with some recent ideas,
as is reflected in the title. The current status of Starobinsky inflation is reviewed and compared to three main conjectures in the Swampland
program. It is argued that the Starobinsky inflation model is not in conflict with those Swampland conjectures in their basic versions.

\newpage

\section{Introduction}

Professor Vladislav Gavrilovich Bagrov was Chair in Quantum Field Theory at Tomsk State University in Western Siberia since 1974.
His influence on students in theoretical physics in Tomsk was immense. I was one of those students in late 70's. Professor Bagrov had great and unique personality, and always defended academic freedom for his students. In his terms, the academic freedom meant 
"do not interfere" and "do not exploit" students - they will find a way in research themselves. At the same time, students were required to observe the rule "the Chair is always right". And it worked! Many of us went to research areas totally unknown in Tomsk. For instance, 
in late 80's, I wrote the first book in Russian about superstring theory \cite{mybook}. 

Life in Siberia was always tough, and surviving there meant taking care of the necessities at all times. A purely theoretical research in Siberia was nonsense for many people, or just fun (for us). Nevertheless, we did it, and for some of us it became a lifetime challenge
\cite{myr}.

In this paper, I want to continue the tradition mentioned above, and briefly address the subject of Starobinsky inflation in connection to the so-called Swampland conjectures originated in string theory when it met cosmology and inflation. Professor Alexei Alexandrovich Starobinsky was a pioneer of cosmological inflation in early 80's, who passed away almost simultaneously with Professor Bagrov. I had a privilege to work with Professor Starobinsky also, though much later \cite{sk1,sk2}. 

This paper is neither a review (there are many reviews about the subjects in the title), nor a historical (or chronological) description, so
that many original contributions are not mentioned or referred to, in order to save space. 

\section{Starobinsky model of inflation}

Cosmological inflation is a proposal (sometimes called a cosmological paradigm) about the existence of a "short" but "fast" 
(exponential, or de-Sitter-type) accelerated grow of the scale factor $a(t)$  in the very early Universe between $10^{-36}$ s and
$10^{-32}$ s, before particle production (reheating) and before the radiation-dominated era described by the Friedman metric.  Inflation is often defined by the equation
\be
\ddot{a}(t)>0 \quad {\rm or, ~equivalently,} \quad \fracmm{d}{dt}(aH)^{-1}<0~,
\ee
where $H(t)$ is the Hubble function, $H=\dot{a}/a$, and the dots denote the time derivatives. Though the functions $a$ and $\dot{a}$ were increasing with time during inflation, the Hubble (or particle) horizon $(aH)^{-1}$, describing the causally connected region in space, was decreasing. Unlike dark energy (accelerating Universe), there was a quick "Graceful Exit" after inflation. An inflationary solution is supposed to be an attractor, in order to eliminate strong dependence upon initial conditions.

There is the significant (indirect) evidence for inflation due to (i) correct predictions of fluctuations and anisotropy of the cosmic microwave
background (CMB) radiation, (ii) explaining the origin of the Friedman universe by solving its internal problems of flatness and horizon, and
the absence of exotic species and heavy relics. Furthermore, (iii) inflation can explain the origin of structure in the current Universe, because it amplified quantum fluctuations that can be seeds of the structure formation.

There are many inflation models consistent with CMB observations because the CMB offers merely a small window to inflation. Any approach needs a driver for inflation, while it is usually taken to be a neutral scalar field called inflaton. When assuming no other particles during inflation, gravity and inflaton would be the only essential players, whose unification leads to the gravitational origin of inflation. Even
without such unification, the standard approach to inflation (known as quintessence) can be reformulated in terms of gravity only, which appears to be a good discrimination tool for all "single-field" inflation models.

The Starobinsky model of inflation \cite{star} can be defined as the generally-covariant and non-perturbative extension of the standard Einstein-Hilbert (EH) gravity theory by the term quadratic in the Ricci scalar curvature $R$. All terms of the higher-order in the spacetime curvature 
are known to be irrelevant in the Solar system, while they were also negligible during reheating after inflation, i.e. in the weak-gravity regime. However, it was not the case during inflation, when the scale of inflation was much higher.

The Starobinsky model can be considered as the particular case of modified gravity. A modified gravity action has the higher-derivatives and generically suffers from {Ostrogradsky instability and ghosts}. However, there are exceptions. In the most general modified gravity action, whose Lagrangian is quadratic in the spacetime curvature, the  only ghost-free term is just given by $R^2$ with a positive coefficient, which leads  to the Starobinsky model with the action 
\be  S= \fracmm{M^2_{\rm Pl}}{2}\int \mathrm{d}^4x\sqrt{-g} \left( R +\fracmm{1}{6M^2}R^2\right)=
\fracmm{M^2_{\rm Pl}}{2}\int \mathrm{d}^4x\sqrt{-g} ~F(R)~,
\ee
having the only (mass) parameter $M$, where $M_{\rm Pl}=1/\sqrt{8\p G_{\rm N}}\approx 2.4\times 10^{18}$ GeV, the spacetime signature is  $(-,+,+,+,)$ and the natural units are used, $\hbar=c=1$.

The metric of a flat Friedman universe is given by 
\be ds^2=-dt^2+a^2\left(dx_1^2+dx_2^2+dx_3^2\right)~.
\ee
Then the action (2) leads to equations of motion in the form
\be  
2H\ddot{H} - \left(\dot{H}\right)^2 + H^2\left(6\dot{H} + M^2\right)=0~,
\ee
When searching for a solution in the form of left Painlev\'e series, $H(t)=\sum^{k=p}_{k=-\infty}c_k(t_0-t)^k$, one finds \cite{kpv}
\be
\begin{split}
H(t)  & =  \fracmm{M^2}{6}(t_0-t)+\fracmm{1}{6(t_0-t)} - \fracmm{4}{9M^2(t_0-t)^3}+
\fracmm{146}{45M^4(t_0-t)^5}  \\
& {} -\fracmm{11752}{315 M^6 (t_0-t)^7} + {\cal O} \left((t_0-t)^{-9}\right)
\end{split}
\ee
valid for $M(t_0-t)>1$. This special solution is an attractor, while $R=12H^2+6\dot{H}$.

In the high-curvature regime relevant for inflation, the EH term can be ignored and the action (2) with only the $R^2$-term becomes 
scale-invariant. In the slow-roll (SR) approximation, $\abs{\ddot{H}} \ll \abs{H\dot{H}}$ and  $\abs{\dot{H}} \ll H^2$, one has
\be 
 H(t) \approx \left (\fracmm{M^2}{6}\right) (t_0-t)~.
\ee
The attractor solution spontaneously breaks the scale invariance of the $R^2$-gravity and, therefore, implies the existence of the Nambu-Goldstone boson (scalaron) that is the physical scalar excitation of the higher-derivative gravity in the given approach. It can be revealed by rewriting the Starobinsky action into the more standard (quintessence) form after the field redefinition 
(Legendre-Weyl transform),
\begin{equation} 
  \varphi =  \sqrt{ \fracmm{3}{2}} M_{\rm Pl}\ln F'(\c)   \quad {\rm and}\quad g_{\m\n}\to \fracmm{2}{M^2_{\rm Pl}}F'(\chi) g_{\m\n}~,
  \quad \chi=R~.
  \ee
It yields
\be S[g_{\m\n},\varphi]  = \fracmm{M^2_{\rm Pl}}{2}\int \mathrm{d}^4x\sqrt{-g} R 
 - \int \mathrm{d}^4x \sqrt{-g} \left[ \frac{1}{2}g^{\m\n}\pa_{\m}\varphi\pa_{\n}\varphi
 + V(\varphi)\right]~,
\ee
in terms of the canonical inflaton $\varphi$ with the scalar potential 
\be
V(\varphi) = \fracmm{3}{4} M^2_{\rm Pl}M^2\left[ 1- \exp\left(-\sqrt{\frac{2}{3}}\varphi/M_{\rm Pl}\right)\right]^2~.
\ee
This potential has a plateau (for large values of $\varphi/M_{\rm Pl}$)  that implies an approximate shift symmetry of the inflaton field, as the consequence of the scale invariance of the $R^2$ gravity or of the approximate scale invariance of the action (2) in the large-curvature regime. The potential (8) also has a positive "cosmological constant" given by the first term in (8) and  induced by the $R^2$ term in the action (2), which can be interpreted as the energy driving inflation. The scale of inflation is determined by the parameter $M$ that is identified with the inflaton mass. The universality class is determined by the critical parameter $\sqrt{2/3}$.

The equivalent actions (2) and (7) are usually referred to Jordan frame and Einstein frame, respectively. The approximate shift symmetry
of the potential (8) is the consequence of the approximate scale invariance of the $R^2$ gravity, which requires the presence of the $R^2$ term in any viable model of inflation on the modified $F(R)$-gravity side. It becomes even more transparent by using the inverse transformation from the Einstein frame to Jordan frame, having the parametric form \cite{kw}
\be
R = \left(  \fracmm{\sqrt{6}}{M_{\rm Pl}}
    \fracmm{d V}{d \varphi} + \fracmm{4V}{M^2_{\rm Pl}} \right) e^{ \sqrt{\frac{2}{3}} 
  \varphi/M_{\rm Pl}}~,    \quad F= \left(  \fracmm{\sqrt{6}}{M_{\rm Pl}}
 \fracmm{d V}{d \varphi} + \fracmm{2V}{M^2_{\rm Pl}} \right) e^{ 2 \sqrt{\frac{2}{3}} \varphi/M_{\rm Pl}}~.
\ee
As is clear from these equations, in the SR approximation (chaotic inflation) the first term in the brackets is much less than the second term, which immediately implies $F(R)\sim R^2$.

Up to this point, no input from CMB observations was used besides theoretical (formal) considerations. The fact that  the Starobinsky model (1980) of inflation is in excellent agreement with the  current CMB measurements \cite{planck} can, therefore, be considered as a non-trivial bonus and as an experimental validation of the theoretical model.

A duration of inflation is usually measured by the e-folds number defined by
\be 
N=\int^{t_{\rm end}}_{t_{\rm start}} H(t) dt~.
\ee
One also uses the running e-folds $N(t)$ instead of the running time $t$, as well as  the co-moving wavenumber $k=2\pi/\lambda$ related to $N(t)$ by the equation $d\ln k =-dN$~.

The SR (running) parameters in Einstein frame are defined by
\be
\ve_{\rm sr}(\varphi) = \fracmm{M^2_{\rm Pl}}{2}\left( \fracmm{V'}{V}\right)^2 \quad {\rm and} \quad 
\eta_{\rm sr}(\varphi) = M^2_{\rm Pl} \left( \fracmm{V''}{V}\right)~,
\ee
in terms of the quintessence scalar potential $V$, where the primes denote the derivatives with respect to $\varphi$. In  Jordan frame,
one uses the Hubble flow functions,
\be 
\epsilon_{H} = -\fracmm{\dot{H}}{H^{2}}~,\quad 
	\eta_{H} = \epsilon_{H} - \fracmm{\dot{\epsilon}_{H}}{2\epsilon_{H} H}~~.
\ee

The amplitude of scalar perturbations at the horizon crossing with the pivot scale $k_*=0.05~{\rm Mpc}^{-1}$  is well known from CMB
measurements (WMAP normalization) as
\be 
A_s= \fracmm{V_*^3}{12\p^2 M^6_{\rm Pl}({V_*}')^2}=\fracmm{3M^2}{8\p^2M^2_{\rm Pl}}\sinh^4\left(
\fracmm{\varphi_*}{\sqrt{6}M_{\rm Pl}}\right)\approx 2 \cdot 10^{-9}~,
\ee
where the result of calculation in the Starobinsky model has also been given. The star-subscript refers to the CMB pivot scale.
Equation (13) allows us to determine the mass parameter $M$ in the Starobinsky model together with the scale of inflation, $H_{\rm inf.}$,
as
\be
M\approx 3 \cdot10^{13}~{\rm GeV}\quad {\rm or} \quad  \fracmm{M}{M_{\rm Pl}}\approx 1.3\cdot 10^{-5}~, \quad {\rm and} \quad
H\approx {\cal O}(10^{14})~{\rm GeV}~.
\ee
Having fixed $M$, we get the Starobinsky model without free parameters.

Next, it comes to the crucial check of the model against the dimensionless cosmological tilts of the power spectrum (the scalar tilt $n_s$ and the tensor-to-scalar ratio $r$), whose values are constrained by the CMB measurements \cite{planck} as follows:
\be 
n_s\approx 1+2\eta_{\rm sr} -6\ve_{\rm sr}\approx 0.9649\pm 0.0042~(68\% {\rm CL}) \quad {\rm and} \quad r< 0.032~(95\% {\rm CL})~,
\ee
where we have added the results of calculations in the Starobinsky model, in the first order with respect to the SR parameters.
The Starobinsky inflation model gives 
\be n_s\approx 1- 2/N  \quad {\rm and} \quad r\approx 12/N^2~,
\ee
that comfortably fit the CMB measurements for $N$ between 48 and 64, with the best fit at $N\approx 56$. The corresponding values in
Jordan frame for the beginning and the end of inflation are $M(t_0-t)\approx 2.5$ and $M(t_0-t)\approx 27$, respectively. Equations (16)
are given in the leading orders with respect to $N$, see \cite{sultan,penn} for more precise results.

Excluding $N$ from equations (16) yields a sharp prediction of the Starobinsky model  for the tensor-to-scalar ratio,
\be r\approx 3(1-n_s)^2~.
\ee
Verifying this prediction is one of the major targets of the LiteBIRD, BICEP and Simons Observatory projects in the world.

The Starobinsky inflation does not exclude the higher-order curvature terms in the action (2), though it implies that those terms were
subleading during inflation, being suppressed by the powers of $H^2/M^2_{\rm Pl}\sim 10^{-8}$. The Starobinsky model is sensitive to quantum (UV) corrections because of its high scale and the inflaton field values near the Planck scale during inflation. Therefore, it is important to determine its UV-cutoff $\L_{\rm UV}$ by studying scaling of scattering amplitudes with respect to energy, $E/\L_{\rm UV}$.
A careful calculation yields \cite{hertzberg}
\be 
\Lambda_{\rm UV}=M_{\rm Pl}~.
\ee
Therefore, the predictions of the Starobinsky model for inflation make sense and the model itself can be considered as a trustable effective field theory due to decoupling of heavy modes expected at the Planck scale \cite{dker}.

\section{Swampland conjectures and Starobinsky model}

According to the existing reviews of the Swampland conjectures in the literature \cite{palti,valen}, there are many conjectures and many versions of them. All of them are about consistency with quantum gravity (or UV completion), either in string theory or in general. The
Swampland program is aimed to discriminate between various effective field theories (EFT).  The Starobinsky model can also be considered as an EFT. Unfortunately, not much known for sure about quantum gravity, so that let us confine ourselves to only three 
Swampland conjectures in their simplest versions, and examine whether the Starobinsky model of inflation is consistent with them.

\subsection{No global symmetries in quantum gravity}

This Swampland conjecture is about the absence of global (or rigid) symmetries in any fundamental theory of quantum gravity.  It claims that only local (or gauge) symmetries are possible in quantum gravity. This conjecture is fully in line with fundamental principles of
 General Relativity but may appear unusual for particle physics based on the Standard Model near or below the electro-weak scale.
 
 As regard the Starobinsky model, it has only approximate (global) scale invariance related to the $R^2$ term alone, which is obviously violated by  the EH term and any other possible higher-order term with respect to the spacetime curvature in the gravitational EFT. Therefore, the Starobinsky model does not violate the no-global-symmetry conjecture.

\subsection{Weak-gravity conjecture}

In its simplest version, this conjecture claims that gravity is the weakest force, for example, in comparison to electromagnetic force.

At first sight, the Starobinsky model of inflation contradicts this conjecture because the Starobinsky inflation relies on the gravitational
force as the only force driving inflation in Jordan frame. However, in Einstein frame, Starobinsky inflation is mainly due to the inflaton selfinteractions described by the scalar potential (8), i.e. due to the inflaton force (or inflaton exchange). Therefore, there is no contradiction between the Starobinsky model of inflation and the weak-gravity conjecture.

One may also argue that during inflation there were no sources of electromagnetic force because there were no charged particles yet (they appeared during reheating after inflation).

\subsection{No-de-Sitter conjecture}

This conjecture claims that no de Sitter (dS) spacetime and no eternal inflation are possible in quantum gravity.

Again, at first sight, it appears to be in contradiction to the Starobinsky model of inflation because the potential (8) has the infinite plateau.
However, the infinite plateau in the Starobinsky model (2) is easily destabilized by the higher-order terms with respect to the spacetime
curvature, which are certainly present in any UV-completion of the EFT action (2), see e.g., \cite{kpv,sabk}. A stronger version of the conjecture, excluding all locally flat inflaton potentials \cite{obied} seems to be in conflict with observations \cite{denef}, see also
\cite{ker} for more arguments.

\section{Conclusion}

Our main conclusion is that the Starobinsky model of inflation is not in conflict with the main three Swampland conjectures in their simplest versions, unlike the claim made in \cite{mpp}. 

In the original paper \cite{star} as well as in \cite{mpp}, the Starobinsky model was assumed to originate from quantum effects (renormalization) due to the matter field (loop) contributions to the energy-momentum tensor on the right-hand-side of the effective
(semiclassical) Einstein equations. According to equations (2) and (3) in \cite{star}, see also \cite{duff}, there are several such terms. A contribution from a single quantized matter field comes with a dimensionless coefficient of the order $10^{-4}$, whereas the coefficient at the $R^2$-action in the Starobinsky model of inflation is of the order $10^9$. Hence, one needs about $10^{13}$ quantized matter fields in order to describe Starobinsky inflation. There is no evidence for the existence of such tower of light fields or particles during inflation with a new fundamental scale. Should such scale exist and be close to the Hubble scale of the order $10^{14}$ GeV predicted by  the Starobinsky inflation, it would invalidate the UV-cutoff (19) and lead to strongly coupled gravity beyond computational control.

It is worth mentioning that the Starobinsky model (2) is the best fit for inflation only.  Outside the high-curvature regime relevant for inflation, 
the coefficient at the $R^2$ term should also be changed. This coefficient is expected to be modulated by other fields \cite{kanedak}. Of course, the $R^2$-inflation may also be ruled out either by future measurements of the tensor-to-scalar ratio with the result significantly
different from the prediction (18) or by observing a significant amount of non-Gaussianity.

A string theory derivation of the Starobinsky inflation is unknown, though it is possible to get the effective scalar potential close to (8) in Einstein frame, see e.g., \cite{mpp2,italy}. Any such potential can be converted to Jordan frame by using (9) with the result close but different from (2). On the other hand, the inflaton potential in string theory can be non-perturbatively generated by using D-brane and anti-D-brane interactions \cite{gia}, which imply the presence of a massive abelian gauge field during inflation. Such field naturally arises in the 
Starobinsky-like inflationary models based on supergravity with spontaneous supersymmetry breaking due to the alternative
Fayet-Iliopoulos term and inflaton belonging to a massive vector supermultiplet \cite{yerk}.

Finally, fine-tuning in describing inflation is not necessarily a drawback because inflation was a unique event in the Universe, whose parameters may not be derivable from pure thoughts. The CMB observations teach us a real lesson about inflation and the Starobinsky model certainly captures significant part of the true story that should also be incorporated into quantum gravity theory.

\section*{Acknowledgements} 

The author is grateful to Ignatios Antoniadis, Eugenio Bianchi, Robert Brandenberger, Michael Duff and Gia Dvali for discussions and correspondence.

\noindent This work was supported by Tokyo Metropolitan University, the Japanese Society for Promotion of Science under the grant No.~22K03624, the World Premier International Research Center Initiative (WPI, Japan), the Tomsk Polytechnic University development program Priority-2030-NIP/EB-004-375-2024 and Tomsk State University.

\end{document}